# Core-excited and shape-type resonances in the micro-solvated Uracil: A CASSCF study

Sneha Arora[a] and Achintya Kumar Dutta[a*]

**Abstract:**

Electronic resonances play an important role in electron attachment-induced processes in biomolecules, and their properties can be significantly influenced by the local molecular environment. Here, we investigate the effect of amino acid micro-solvation on the uracil resonances by employing uracil-glycine as a model system. The resonance spectrum of the uracil-glycine complex consists of four π-type shape resonances and three core-excited resonances, including an additional glycine-centered resonance, as characterized using the CASSCF/Resonance via Padé (RVP) methodology. Comparison with isolated uracil and the uracil(ghostGly) model shows that explicit interaction with glycine stabilizes both the shape and core-excited resonances by lowering their energies and increasing their lifetimes, while the ghost calculations demonstrate that basis-set extension alone cannot account for the observed stabilization. The core-excited resonances exhibit states that retain non-negligible lifetimes despite their much higher energy, suggesting that they may play an important role in electron-induced dissociation pathways.  Overall, the present results demonstrate that amino acid micro-solvation significantly modifies the resonance landscape of uracil, highlighting the importance of explicitly accounting for local biomolecular interactions in theoretical studies of electron attachment.

*achintya@chem.iitb.ac.in

## 1. Introduction:

Resonances are short-lived metastable electronic states that play a fundamental role in wide range of physical, chemical, and biological processes.[1–10] These are encountered in diverse areas, ranging from plasma physics and ultrafast spectroscopy to interstellar chemistry and radiation-induced chemistry. In biological systems, low-energy electrons (LEEs) produced during the interaction of ionizing radiation with matter can temporarily attach to biomolecules, forming transient negative ions (TNIs) or resonance states. These metastable anions have finite lifetimes, typically spanning from a few femtoseconds to several picoseconds, during which the excess electron may either undergo autodetachment or dissociative electron attachment (DEA), leading to bond cleavage and molecular fragmentation.[1–3,5–11] Therefore, investigating the energies, lifetimes, and decay dynamics of electronic resonances is essential for understanding electron-induced processes in molecular systems.

Depending on the electronic structure of the target molecule and the energy of the incident electron, electron attachment can give rise to different classes of resonances. In nucleobases, the most important are shape and core-excited/Feshbach resonances. Shape resonances are formed when the incoming electron is temporarily trapped in an unoccupied antibonding orbital, most commonly a $\pi^*$ orbital, of the neutral molecule in its electronic ground state. Although $\sigma^*$-type shape resonances may also occur, they are generally much shorter-lived because of their stronger coupling to the electronic continuum.[9,10,15–20] Core-excited resonances, on the other hand, arise from electron attachment to electronically excited states of the neutral molecule. These states are characterized by a vacancy in an occupied orbital together with two electrons occupying virtual orbitals, giving rise to electronic structures and decay dynamics that differ from those of shape resonances.

Early theoretical and experimental studies suggest that shape resonances facilitate electron transfer from $\pi^*$ orbital of nucleobases to $\sigma^*$ orbital of the C−O or C−N bonds leading to dissociative electron attachment. However, Sanche and co-workers concluded that the short-lived transient negative ion cannot cause C−N bond rupture. It is the long-lived core-excited resonance that leads to this cleavage. Electron transmission experiments have identified resonance features in the 4-6 eV and 9-11 eV regions of the yield functions, which are likely associated with the initial formation of Feshbach or core-excited shape resonances.[21] Core-excited shape resonances in nucleobases occur at energies higher than corresponding neutral excited states, typically above 3-4 eV. These resonances are significant in interactions with electrons above this threshold.[9] However, electron attachment in this energy range remains less well understood due to the greater complexity involved in modelling core-excited resonances theoretically.

Various approaches have been employed to determine the position of the lowest $\pi^*$ shape resonances. We have also calculated the $\pi^*$ shape resonances of the RNA/DNA nucleobases and base pairs using EA-EOM-DLPNO-CCSD/RVP method in our previous works. However, little is known about high energy core-excited/Feshbach resonance states. The core-excited/Feshbach resonance state, denoted as $\pi^1(\sigma*)^2$, which corresponds to electron attachment to $\pi^1(\sigma*)^1$ excited state of uracil, was previously estimated by Kawarai et al.,[22] as they proposed its potential involvement in dissociative electron attachment, though this hypothesis remains unverified. Dora et al.[23] reported 2A′ Feshbach resonances for uracil at energies of 6.17 eV (width: 0.15 eV), 7.62 eV (width: 0.11 eV), and 8.12 eV (width: 0.14 eV); however, the specific electronic configurations of the resonance states have not been identified. Furthermore, Fennimore et al.[9] estimated energy of the core-excited $\pi^1(\pi*)^2$ state of uracil to be within the range of 4.6 to 4.8 eV using EA-EOM-CCSD and complete active space self-consistent field (CASSCF) combined with extended multi-configuration quasi degenerate perturbation theory (XMCQDPT2). The obtained value aligns closely with the experimentally observed peak at 4.6 eV in the energy range of single and double-strand break yields induced by electron transmission on plasmid DNA films. This strong agreement provides compelling evidence that the obtained maxima correspond to core-excited resonance states.

In this study, we employed the Resonance via Padé (RVP) stabilization approach combined with the CASSCF method to compute the resonance states of a micro-solvated uracil model system, specifically uracil(Gly). The primary objective of this manuscript is to use the well-correlated CASSCF method to calculate both the electron-attachment-induced $\pi^*$ shape resonances and the core-excited resonances of the RNA uracil nucleobase within a biological context. To avoid the complexity of modeling the biological environment surrounding RNA, including interactions with histone proteins, we have opted for a simplified model system that incorporates the isolated uracil nucleobase and a glycine amino acid. This approach facilitates study of the resonance states while emphasizing the impact of higher energy electrons capable of accessing core-excited states. While our simplified model effectively captures local electronic interactions, it overlooks critical aspects of the comprehensive biological environment, such as solvent effects, secondary structures, and macromolecular interactions. However, overly detailed models can hinder practical calculations and obscure broader insights. Consequently, our simplified model provides valuable insights into the influence of amino acids on nucleobase-centered resonances in RNA.

## 2. Computational details:

The geometries of isolated and micro-solvated uracil molecules are optimized using the RI-MP2[24,25] with the def2-TZVP[26] basis set, implemented in the ORCA 5.0.3[27,28] and subsequently re-optimized in GAUSSIAN software package at the MP2/cc-pVTZ level with Cs symmetry. Resonance calculations for the molecules are carried out using the CASSCF method in the MOLPRO.[29] The energy stabilization curves are generated by augmenting the cc-pVDZ basis with additional diffuse 1s, 1p, and 1d functions. The cc-pVDZ+1s**1p**1d basis set provides a balance between computational cost and an accurate description of the diffuse resonance states. Since the resonance states investigated are predominantly π-type in character, scaling of the diffuse p functions was found to produce smoother and better-defined stabilization plots than scaling all diffuse functions. Accordingly, this approach was adopted throughout the present study. So, the diffuse 1p functions are selectively scaled by dividing their gaussian exponents by a real scaling factor (α). The choice of active space is guided by the necessity to accurately describe the relevant valence and diffuse virtual orbitals involved in the resonant electron attachment process. The active space for micro-solvated nucleobases is constructed to include all chemically relevant π and π* orbitals, with particular emphasis on incorporating one or more highly diffuse π* orbitals. This choice is essential for enabling the formation of orthogonally discretized continuum states and for accurately describing resonances within the CASSCF framework. A schematic representation of the CASSCF orbital partitioning is shown in Figure 1. As an initial step, shape and core-excited resonances of isolated uracil are computed using a CAS(9,8) active space. Subsequently, to account for micro-solvation effects, the active spaces for the uracil(Gly) and uracil(ghostGly) complexes are defined as CAS(13,12) and CAS(9,8), respectively. Here, CAS(N,M) denotes an active space comprising N active electrons distributed among M active orbitals. Figure 3 presents the stabilization plot for the uracil obtained at the CASSCF level of theory.

Resonance energies and widths were extracted from the stabilization data using the RVP method.[30–33] Consequently, in a stabilization graph, resonance states appear as stable energy regions with respect to the scaling parameter α, whereas continuum states display significant energy fluctuations as α varies. The RVP calculations are conducted using the open-source software "Automatic RVP."[30,34]

## 3. Results and Discussion:

In this section, we will discuss the CASSCF/RVP results obtained for micro-solvated uracil systems.

### 3.1 Uracil and Glycine resonances:

Before investigating the influence of glycine micro-solvation on the resonance properties of uracil, we first examine the isolated uracil and glycine molecules. The resonance characteristics of the individual constituents provides an important reference for understanding the changes introduced upon complex formation. Since the resonance spectrum of isolated uracil has been thoroughly investigated in previous theoretical studies and its low-lying shape resonances are well supported by experiment, it serves as an ideal benchmark for validating the present CASSCF/RVP calculations. The resonance characteristics of isolated uracil were investigated using the CASSCF/RVP approach with a CAS(9,8) active space. This CAS(9,8) active space was selected to include the frontier π and π* molecular orbitals that are directly involved in the low-lying excited states and electron-attachment resonance states of uracil. This choice provides a balanced description of the occupied and virtual π orbitals relevant to the resonance processes investigated in this work. A total of three π*-type shape resonances and two core-excited resonances were identified. The corresponding active space orbitals, stabilization plot, and natural orbitals are presented in Figures 2 and 3, and the calculated resonance energies and widths are summarized in Table 1. To facilitate direct comparison with previous theoretical studies and the available experimental data, the resonance energies were scaled with respect to the experimental 1π* resonance energy of isolated uracil (0.22 eV). The resulting resonance spectrum provides the reference framework for assessing the changes induced by glycine micro-solvation, which are discussed in the following sections.

After scaling with the experimental $1\pi^*$ resonance energy, the lowest $1\pi^*$ shape resonance is obtained at 0.22 eV with a lifetime of approximately 12 fs, corresponding to the longest-lived state of the three shape resonances. The associated natural orbital shows that the excess electron occupies the lowest antibonding $\pi^*$ orbital localized over the uracil nucleobase. The second shape resonance, $2\pi^*$, is identified at 1.82 eV and exhibits a comparable lifetime of approximately 11 fs. Similar to the $1\pi^*$ state, the excess electron remains predominantly localized on the nucleobase. The highest-lying $3\pi^*$ resonance is observed at 5.06 eV and possesses the shortest lifetime of approximately 9 fs among the three shape resonances. The systematic decrease in lifetime from the $1\pi^*$ to the $3\pi^*$ state reflects the increasing coupling of the higher-energy resonances to the electronic continuum, which enhances electron autodetachment and consequently reduces the lifetime of the transient anion.

In addition to the three $\pi^*$-type shape resonances, two core-excited resonances are identified at 5.60 and 10.42 eV, corresponding to the $\pi^1(\pi^1)^*$ and $\pi^1(\pi^2)^*$ configurations, respectively. Analysis of the corresponding natural orbitals indicates that both resonances arise from excitation of an inner-valence π electron into a π* orbital in the presence of the attached electron, thereby exhibiting the characteristic Feshbach-type electronic character. The lower-lying $\pi^1(\pi^1)^*$ core-excited resonance has a lifetime of approximately 8 fs, which is comparable to that of the $3\pi^*$ shape resonance despite occurring at a higher energy. The second core-excited resonance, $\pi^1(\pi^2)^*$, is located at 10.42 eV and exhibits a slightly shorter lifetime of approximately 7 fs. Although these resonances occur at higher energies than the shape resonances, they remain relatively long-lived. This behavior is characteristic of Feshbach resonances, whose decay requires electron-correlation-mediated rearrangement before electron emission can occur, making their autodetachment less efficient than that of shape resonances.

The present resonance energies are first compared with the multiconfigurational CASSCF results reported by Fennimore and Matsika[9] using a CAS(11,9)/6-31G active space. The $1\pi^*$ and $2\pi^*$ shape resonances show very good agreement with their calculations, while the $3\pi^*$ resonance is higher in energy by approximately 0.2 eV. For the core-excited resonances, the first $\pi^1(\pi^1)^*$ state is obtained at 5.60 eV, in excellent agreement with the value reported by Fennimore and Matsika[9]. No literature value is available for the higher lying $\pi^1(\pi^2)^*$ resonance state, however, exhibits a larger deviation, which is expected given the increased sensitivity of higher-lying core-excited resonances to the choice of active space and the treatment of electron correlation. The resonance widths also differ between the two studies, reflecting the greater sensitivity of resonance lifetimes to the underlying theoretical methodology. We also compared these results with our previous EA-EOM-DLPNO-CCSD/RVP calculations performed with the cc-pVDZ+2s2p2d basis set.[35] For a consistent comparison, the reported resonance energies were also scaled with respect to the experimental 1π* resonance energy. The 2π* and 3π* resonances are higher by approximately 0.3 and 0.6 eV, respectively, while the width of the 1π* resonance increases by about 0.04 eV relative to the EA-EOM-DLPNO-CCSD results. In contrast, width of the 3π* resonance is somewhat reduced, whereas the 2π* width is slightly larger than the corresponding EA-EOM-DLPNO-CCSD value. Comparison with the available experimental data[36] shows that the overestimation of the third π* resonance has been consistently observed in previous theoretical studies. Overall, the present CASSCF/RVP calculations reproduce the ordering and character of the resonance states reported in earlier theoretical investigations, while the remaining quantitative differences are largely confined to the higher-lying resonances and the resonance widths.

Before examining the effect of glycine micro-solvation on uracil, it is important to study and characterize the resonance states in the glycine anion. The resonance calculations for glycine are performed at CAS(5,5)/cc-pVDZ+1s**1p**1d level. The corresponding active space orbitals are shown in Figure 4, while the calculated resonance parameters are summarized in Table 1. Analysis of the stabilization plot reveals the presence of a single π*-type shape resonance. For consistency with the uracil calculations, the resonance energy was calibrated with respect to the experimentally reported 1π* resonance energy of glycine (1.93 eV). The calculated resonance width is approximately 0.073 eV, corresponding to a lifetime comparable to that of the 2π* shape resonances of uracil. The associated natural orbital analysis shows that the excess electron is localized predominantly on the glycine moiety, confirming the π*-type character of the resonance. These calculations establish the reference resonance properties of isolated glycine, which serve as the basis for understanding the changes observed upon formation of the uracil-glycine complex.

### 3.2 Shape and core-excited resonances in uracil(Gly):

#### 3.2.1 Shape resonances:

To investigate how a biologically relevant molecular environment influences transient anion formation beyond simple aqueous solvation, we extended our resonance study to the micro-solvated uracil-glycine model. Glycine was selected as the simplest amino acid model because it provides a realistic representation of local protein interactions while maintaining a computationally tractable system. An amino acid possesses additional valence orbitals capable of participating directly in the electronic structure of transient anionic states. Consequently, the uracil(Gly) complex provides an ideal model for probing the influence of specific biomolecular interactions on both shape and core-excited resonances of uracil.

The resonance spectrum of the uracil(Gly) complex was characterized using the multiconfigurational CASSCF/RVP methodology with a CAS(13,12) active space. The selected active space includes all molecular orbitals directly involved in the electron-attachment process and adequately describes both the low-lying shape resonances and the higher-lying core-excited configurations. The corresponding active-space orbitals are presented in Figure 5. The resonance parameters were extracted from stabilization graphs obtained at the RVP/CAS(13,12)/cc-pVDZ+1s1p1d level of theory (Figure 6). Following the procedure adopted throughout this work, the calculated resonance energies were referenced to the experimentally established position of the lowest uracil ($1\pi^*$) resonance (0.22 eV), thereby allowing a direct comparison with available experimental and theoretical studies. The resulting resonance energies and widths are summarized in Table 2.

The calculated spectrum consists of four $\pi^*$ shape resonances together with three core-excited resonances. The overall number and ordering of the shape resonances closely resemble those obtained previously using the EA-EOM-DLPNO-CCSD/RVP methodology for isolated uracil and the uracil-glycine complex, demonstrating that the present multiconfigurational treatment successfully reproduces the essential electronic structure of the transient anion while simultaneously enabling an explicit description of higher-lying core-excited states.[38] The qualitative agreement between the two theoretical approaches supports the reliability of the RVP methodology for characterizing metastable electronic states in biologically relevant molecular systems.

The lowest shape resonance, ($1\pi^*$), is located at 0.22 eV and remains strongly localized on the uracil nucleobase. The corresponding natural orbital (Figure 6) shows that the excess electron occupies the lowest antibonding $\pi^*$ orbital of uracil, indicating that the fundamental character of the lowest resonance is preserved despite the presence of the neighboring amino acid. The calculated resonance width corresponds to a lifetime of approximately 15 fs, indicating that although the electron remains metastable, the amino acid environment provides additional stabilization relative to isolated uracil. This observation is fully consistent with our previous EA-EOM-DLPNO-CCSD/RVP investigation, in which glycine was found to stabilize the nucleobase-centered resonances more effectively than micro-hydration by water molecule.[38] In that study, the presence of glycine significantly increased the lifetime of the $1\pi^*$ resonance while simultaneously lowering its energy, demonstrating that specific amino acid interactions can alter resonance states more efficiently than a purely aqueous environment.

The next state, the $2\pi^*$ resonance, is calculated at 1.44 eV and also remains predominantly localized on the uracil moiety. The corresponding natural orbital exhibits a small but distinct contribution from the glycine moiety, indicating the weak intermolecular electronic coupling between the nucleobase and the amino acid. Although the excess electron remains primarily associated with the nucleobase, the neighboring amino acid perturbs the electronic distribution sufficiently to introduce partial delocalization across the intermolecular interface. Such behavior is characteristic of weak orbital mixing between adjacent molecular fragments and suggests that the local protein environment can influence metastable electron attachment even when the excess electron is predominantly localized on the nucleobase. Similar stabilization of the low-lying shape resonances was observed in our previous micro-solvation study.[38] A particularly interesting feature of the present calculations is the appearance of an additional resonance centered predominantly on the glycine fragment. This state, denoted as ($\pi^*$(glycine)), occurs at 2.91 eV with a width of 0.065 eV and was absent in isolated uracil. Natural orbital analysis reveals that the excess electron is predominantly localized on the glycine moiety, demonstrating that the amino acid provides an additional acceptor orbital that directly participates in the metastable anionic electronic structure following electron attachment. Consequently, the excess electron is no longer restricted to nucleobase-centered orbitals but may instead be localized on different molecular fragments depending on the electronic coupling between the two constituents. This observation also considerably extends the conclusions of our previous EA-EOM-DLPNO-CCSD/RVP study, thereby confirming that the appearance of glycine-centered resonance state is an intrinsic feature of the electronic structure rather than an artifact of the computational methodology.

The highest-lying shape resonance, ($3\pi^*$), is located at 5.04 eV and exhibits the shortest lifetime among the observed shape resonances, approximately 9 fs. The increased resonance width reflects stronger coupling with the electronic continuum and consequently more rapid autodetachment of the excess electron. Despite its relatively short lifetime, this state is of particular importance because it lies energetically close to the lowest core-excited resonance discussed below. Such energetic proximity raises the possibility of configurational interaction between shape and core-excited states, an effect that has also been emphasized in previous theoretical investigations of isolated uracil.[9] The energetic proximity of these resonances is expected to enhance configurational mixing, which may alter the decay dynamics of the resonance states and consequently influence subsequent electron-attachment induced processes.

The calculated shape resonances reveal that the amino acid environment influences resonant electron attachment through two complementary mechanisms. First, it stabilizes the conventional nucleobase-centered shape resonances by modifying the local electrostatic and orbital environment. Second, and more importantly, it introduces new electron-attachment channels localized on the amino acid itself. These findings demonstrate that specific amino acid-nucleobase interactions cannot be regarded merely as a perturbation of isolated nucleobase resonances but instead lead to qualitative changes in the electronic structure of resonance anions. Such behavior has important implications for understanding electron-induced chemistry in biologically realistic environments, where proteins and nucleic acids coexist in close proximity. Furthermore, the appearance of a glycine-centered resonance in the simplest micro-solvated complex suggests that increasing the number of surrounding amino acids may give rise to additional intermolecular resonance states involving multiple competing electron-localization pathways. Although the present work considers only a single glycine molecule, these results motivate future investigations of larger micro-solvated clusters and bulk environments, where intermolecular interactions may further diversify the metastable anion landscape and ultimately influence the efficiency of low-energy electron-attachment induced damage in biological systems.

**3.2.2: Core-excited Resonances:**

In addition to the four $\pi^*$ shape resonances, the present CASSCF/RVP calculations identify three distinct core-excited (Feshbach) resonances in the micro-solvated uracil–glycine complex (Table 2). Unlike shape resonances, which originate from direct occupation of an unoccupied molecular orbital by the incoming electron, core-excited resonances arise from a two-electron excitation process in which an electron from an occupied orbital is simultaneously promoted to a virtual orbital while the incident electron occupies another bound orbital. The resulting electronic configuration consists of an excited neutral core together with an additional bound electron, giving rise to a metastable anionic state that is energetically embedded within the ionization continuum. Because of their two-electron excited character, these states exhibit characteristic resonance behavior, where electron detachment requires a multielectron rearrangement rather than simple one-electron autodetachment. Consequently, core-excited resonances generally possess longer lifetimes than shape resonances despite occurring at considerably higher energies.

The lowest core-excited resonance, denoted as $\pi^1(\pi^1)^*$, is located at 5.55 eV and exhibits a lifetime of approximately 7 fs. Natural orbital analysis (Figure 7) demonstrates that this state originates from promotion of an electron from an inner valence π orbital of uracil into the lowest unoccupied 1π orbital while the incoming electron occupies the resulting anionic configuration. Interestingly, this resonance lies only about 0.5 eV above the highest shape resonance ($3\pi^*$, 5.04 eV), placing the two states in close energetic proximity. Previous theoretical investigations of isolated uracil have suggested that such neighboring shape and core-excited resonances may exhibit configurational interaction or electronic mixing because of their similar energies.[9] The present calculations indicate that this feature is retained in the uracil-glycine complex, suggesting that the energetic relationship between the high-lying shape resonance and the lowest core-excited resonance remains essentially unchanged despite the presence of the amino acid.

A second core-excited resonance, $\pi^1(\pi^2)^*$, is identified at 8.60 eV with a lifetime of approximately 6 fs. This state corresponds to excitation from the same inner π orbital into the higher lying $2\pi^*$ orbital of uracil. Similar to $\pi^1(\pi^1)^*$, both the hole and the excited electron remain localized predominantly on the nucleobase. An important consequence of complexation with glycine is the appearance of an additional core-excited resonance, $\pi^2(\pi^g)^*$, at 8.27 eV, which is energetically stabilized by approximately 0.33 eV relative to the analogous uracil-centered $\pi^1(\pi^2)^*$ state. Natural orbital analysis reveals that this resonance differs qualitatively from the two lower core-excited states. Rather than promoting an electron into another antibonding orbital localized on uracil, the excitation transfers electron density into an acceptor orbital predominantly localized on the glycine fragment. Consequently, the excited electronic configuration is no longer confined to the nucleobase but instead becomes intermolecular in nature, with the hole remaining on uracil while the excited electron occupies an orbital associated primarily with the amino acid. Such an excitation pathway has no analogue in isolated uracil and represents a direct consequence of introducing a chemically active molecular environment capable of participating in the transient electronic structure.

A comparison with previous theoretical studies further emphasizes the significance of these findings. Matsika and co-workers demonstrated that the resonance spectrum of isolated uracil is dominated by nucleobase-centered $\pi \rightarrow \pi^*$ excitations, with the lowest Feshbach resonances retaining a purely intramolecular character despite the presence of significant configurational mixing with nearby shape resonances. The present calculations reproduce this behavior for the $\pi^1(\pi^1)^*$ and $\pi^1(\pi^2)^*$ states, thereby confirming that the essential electronic structure of the nucleobase remains remarkably robust upon micro-solvation. At the same time, however, the uracil-glycine complex exhibits an entirely new excitation channel in which the excess electron is transferred from the nucleobase to the amino acid fragment. The coexistence of these intramolecular and intermolecular excitation pathways represents the principal difference between isolated uracil and the micro-solvated complex and illustrates how specific amino acid interactions enrich the resonance landscape beyond that accessible in the isolated molecule. Although the lowest Feshbach resonances continue to resemble those of isolated uracil, the amino acid environment introduces an additional intermolecular core-excited state in which the excited electron occupies a glycine-centered orbital. The similarity between the behavior of the shape and core-excited resonances suggests that the role of amino acids extends well beyond simple electrostatic stabilization. Instead, amino acids act as electronically active components of the local environment, capable of modifying both electron attachment and subsequent electronic excitation processes. Although the present study considers only a single glycine molecule, these findings imply that progressively larger amino acid environments may support an increasingly rich manifold of intermolecular resonance states arising from electronic coupling between multiple neighboring molecules.

Despite lying at considerably higher energies than the shape resonances, the core-excited states retain non-negligible lifetimes ($\approx$ 3.5-7 fs); the lowest-lying $\pi^1(\pi^1)^*$ state approaches the lifetime of the shortest-lived shape resonance ($3\pi^*$, $\approx$ 9 fs), whereas the higher-lying $\pi^1(\pi^2)^*$ and $\pi^2(\pi^g)^*$ states are in fact, shorter-lived than all four shape resonances. This behavior is a direct consequence of their Feshbach character. Whereas shape resonances decay predominantly through rapid one-electron autodetachment into the continuum, decay of a core-excited resonance requires electron-correlation driven rearrangement of the excited electronic configuration before electron emission can occur. The longer lifetime provides a larger time window for nuclear relaxation to occur while the electron remains temporarily bound, allowing bond stretching, angle deformation, and vibrational energy redistribution to develop on the anionic potential energy surface. Consequently, core-excited resonances are expected to play a particularly important role in facilitating dissociative electron attachment.

**3.3 uracil(ghostGly):**

To distinguish the effects of genuine intermolecular interactions from those arising solely due to the presence of additional basis functions, we performed an additional set of calculations in which glycine was treated as a ghost fragment. In this approach, only the basis functions of glycine were retained, thereby eliminating any direct electronic interaction with uracil. Consequently, comparison of the resonance properties of uracil(ghostGly) with those of the fully interacting uracil(Gly) complex enables the changes arising from the basis set extension to be separated from those originating from the physical presence of the amino acid. The uracil(ghostGly) calculations were carried out using the same CAS(9,8) active space employed for the isolated uracil anion. This active space corresponds to the uracil-centered orbitals contained within the larger CAS(13,12) active space used for the uracil(Gly) calculations, ensuring a consistent description of the uracil resonance states and allowing a direct comparison between the isolated, ghost, and fully interacting systems. The corresponding stabilization plot for uracil(ghostGly) is shown in Figure 6.

After applying the scaling procedure with respect to the experimental lowest $\pi^*$ resonance energy of 0.22 eV for uracil, we identified three shape resonances in the uracil(ghostGly) system, consistent with the isolated uracil calculations (Table 2). The glycine-centered resonance is absent in this model because the glycine fragment is treated as a ghost, thereby eliminating the possibility of charge localization on the amino acid. A comparison of the resonance energies between uracil(ghostGly) and the micro-solvated uracil(Gly) system reveals notable differences. The $2\pi^*$ resonance is located at 2.10 eV, approximately 0.7 eV higher in energy than in uracil(Gly), while the $3\pi^*$ resonance is shifted upward by about 0.2 eV. In addition, the width of the 3π* resonance increases by approximately 0.03 eV, indicating a corresponding reduction in its lifetime. A similar trend is observed for the remaining shape resonances, all of which exhibit broader widths than their counterparts in the interacting uracil-glycine complex. In addition to energy shifts, the lifetimes of all shape resonances in the ghost model are significantly decreased, as reflected in their broader widths compared to those observed in uracil(Gly). This decrease in lifetime highlights the crucial role of glycine in providing stabilization to the resonance states.

We have also identified two core-excited resonances, $\pi^1(\pi^1)^*$ and $\pi^1(\pi^2)^*$. The lower-lying $\pi^1(\pi^1)^*$ resonance is located close in energy to the 3π* shape resonance and exhibits a lifetime of approximately 5 fs. Natural orbital analysis shows that this state originates from excitation of an inner-valence π electron into the lowest unoccupied 1π* orbital of uracil, giving rise to the characteristic two-electron excited electronic configuration. Similar to the isolated uracil and uracil(Gly) systems, the energetic proximity of the highest shape resonance and the lowest core-excited resonance is preserved in the ghost model, indicating that this feature is an intrinsic characteristic of the uracil resonances. The second core-excited resonance, $\pi^1(\pi^2)^*$, is identified at 10.85 eV with a lifetime of approximately 4 fs. Its lifetime is reduced by about 1 fs compared to uracil(Gly). These results indicate that the absence of explicit glycine interaction leads to a noticeable destabilization of the core-excited resonance states.

Overall, the uracil(ghostGly) calculations serve as an important control for assessing the origin of the resonance stabilization observed in the fully interacting uracil(Gly) complex. Although the additional basis functions provided by the ghost fragment produce measurable changes in the resonance energies, they do not reproduce either the magnitude of stabilization or the appearance of the glycine-centered resonance observed in the interacting uracil-glycine system. These results demonstrate that the stabilization of both the shape and core-excited resonances arises predominantly from the explicit interaction between uracil and glycine rather than from basis-set effects alone.

## 4. Conclusions:

Our study shows that the interaction of glycine with uracil plays a key role in stabilizing both shape and core-excited resonances of the nucleobase. In the micro-solvated uracil(Gly) system, we identified four $\pi^*$ shape resonances and three core-excited resonances, with glycine not only introducing a glycine-centered resonance but also lowering the energies and extending the lifetimes of uracil-centered states. In contrast, when glycine was replaced by a ghost atom, the resonance states were found to be higher in energy, broader in width, and therefore shorter-lived, clearly indicating reduced stability in the absence of explicit interactions. The core-excited states, although higher in energy, have lifetimes comparable with the high-lying shape resonances, suggesting their potential importance in electron-driven dissociation processes. Together, these results highlight how micro-solvation effects, such as orbital mixing and electrostatic stabilization, can significantly influence the resonance landscape of nucleobases, with direct implications for understanding radiation-induced processes in biological environments.

**Tables and Figures:**

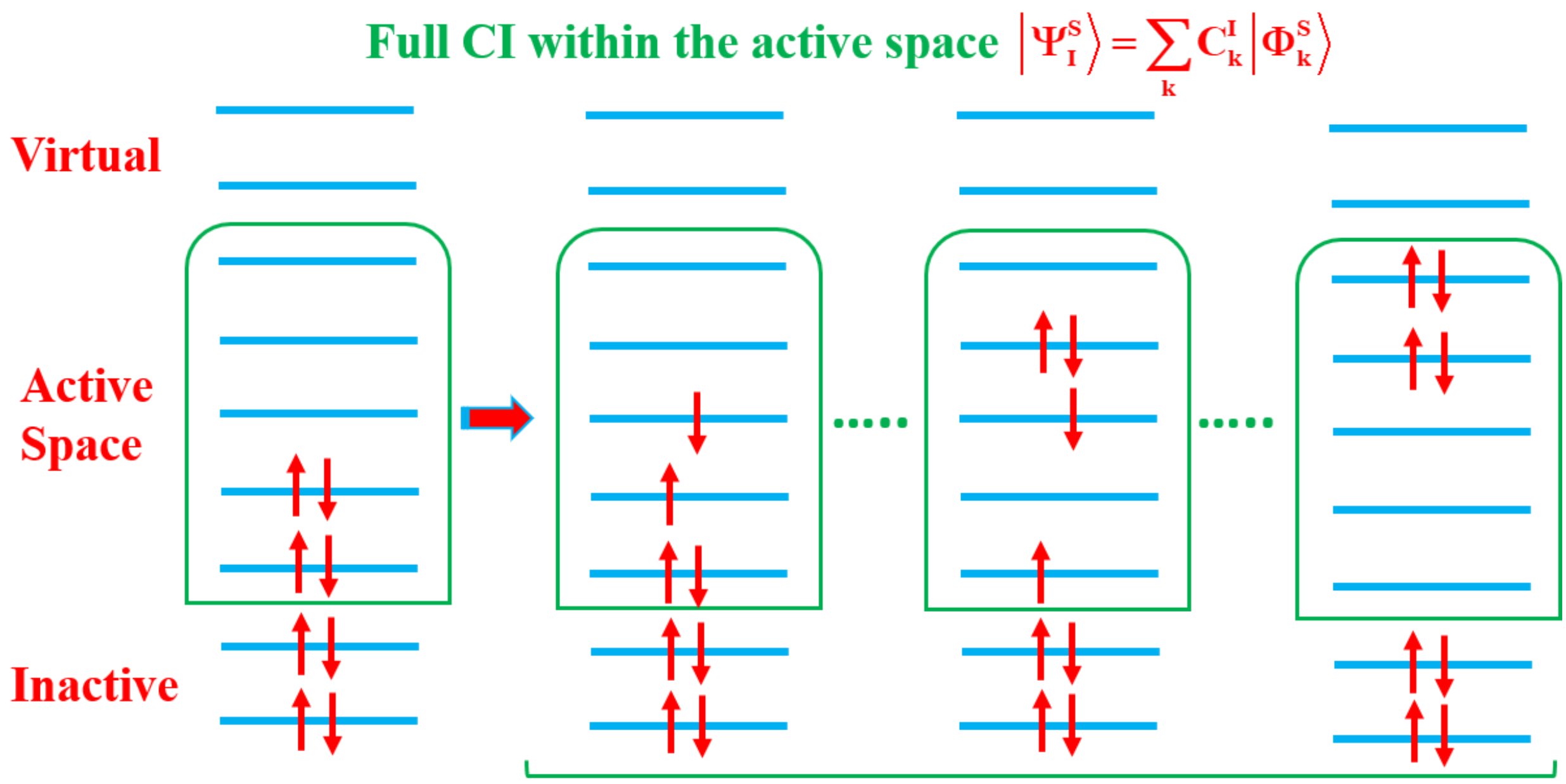

*Figure 1. Schematic of the CASSCF orbital partitioning into inactive (doubly occupied), active (variationally correlated), and virtual (unoccupied) subspaces.*

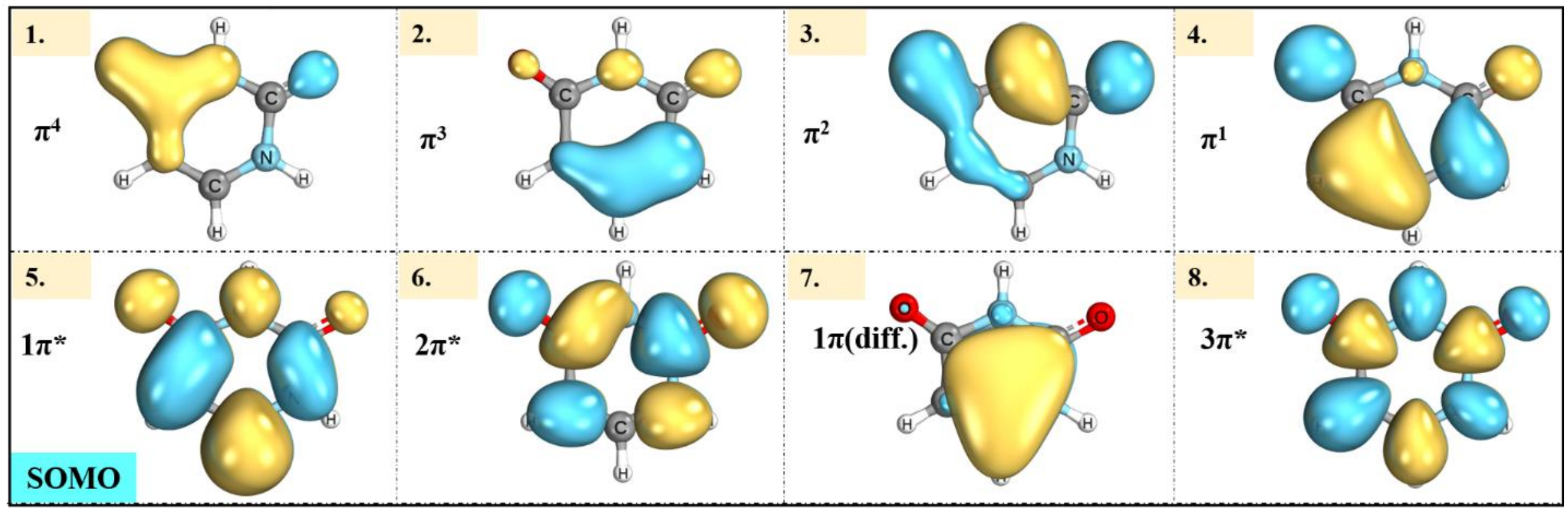

*Figure 2. The CAS(9,8) active space orbitals for isolated uracil anion.*

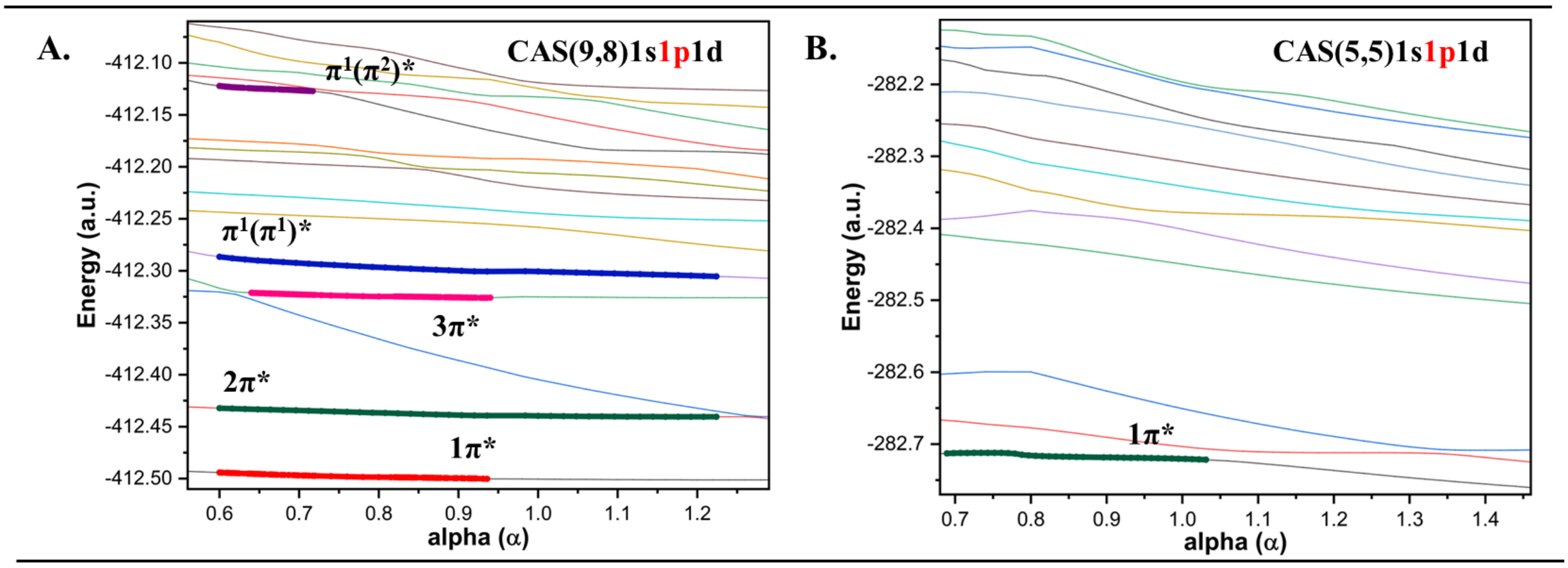

*Figure 3. The energy stabilization graphs for isolated A. uracil and B. glycine anions with the highlighted resonances states.*

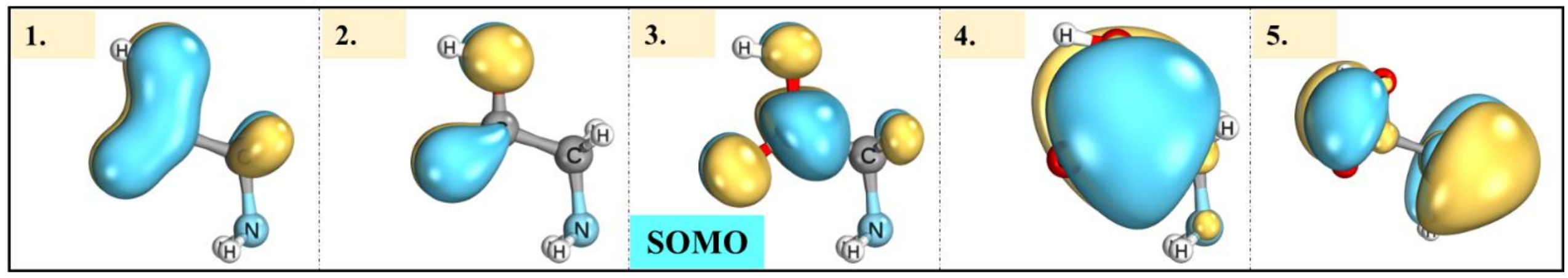

*Figure 4. Glycine CAS(5,5) active space orbitals.*

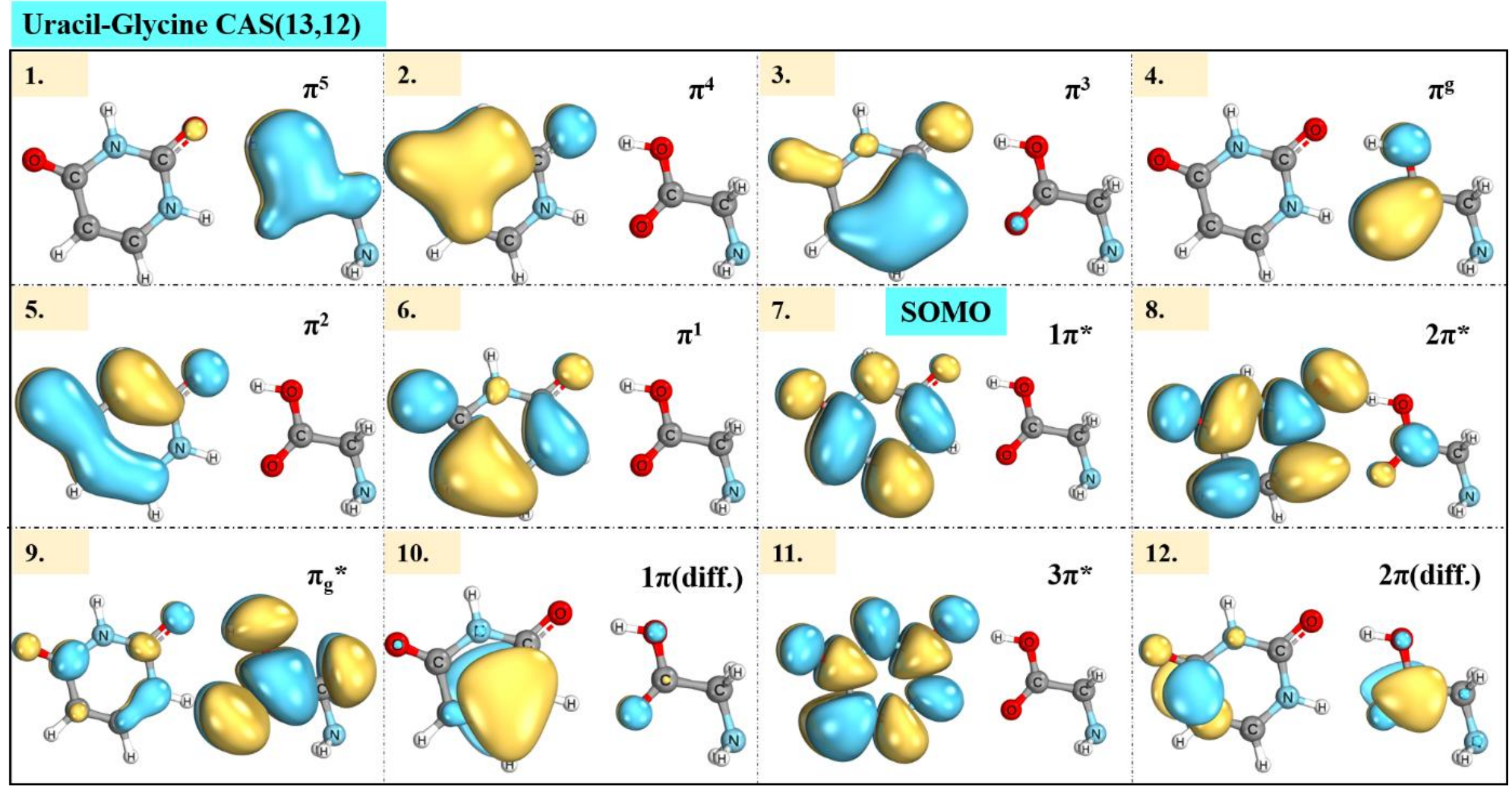

*Figure 5. CAS(13,12) active space orbitals for uracil glycine model (uracil(Gly)) in cc-pVDZ+1s**1p**1d basis.*

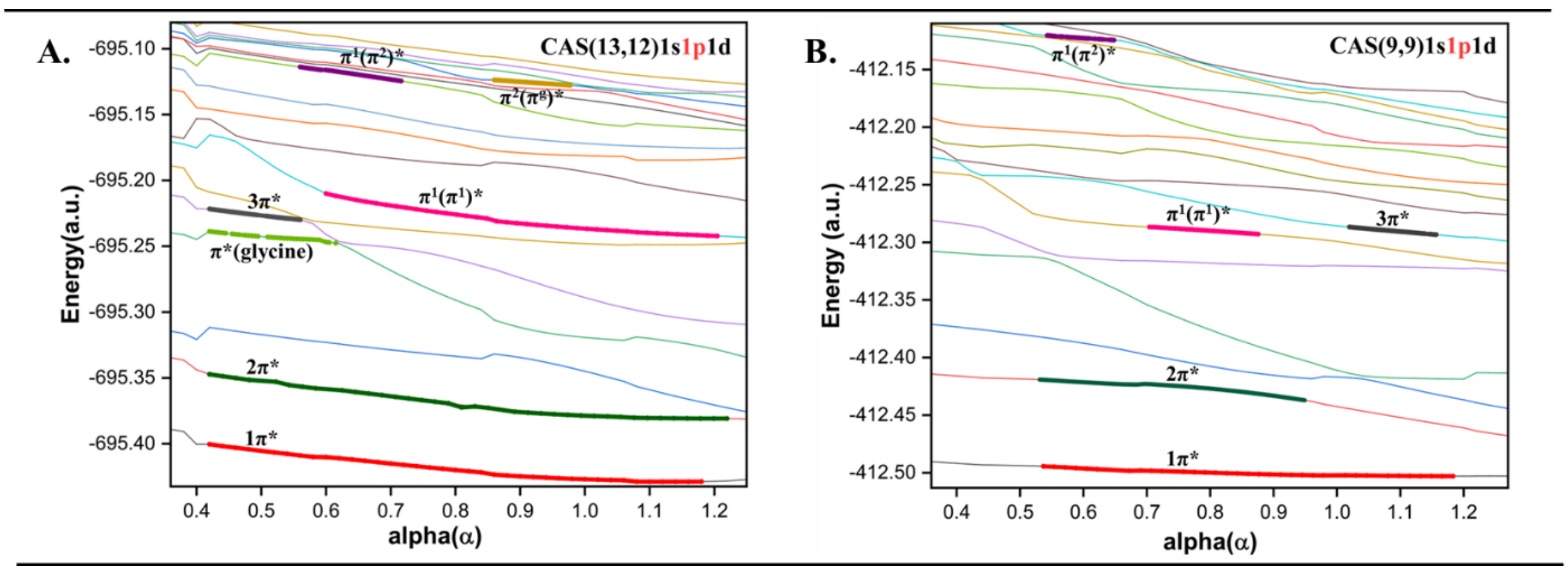

*Figure 6. Stabilization plots of A.) uracil(Gly) and B.) uracil(ghostGly) with the highlighted shape and core-excited resonances at CASSCF/RVP level of theory.*

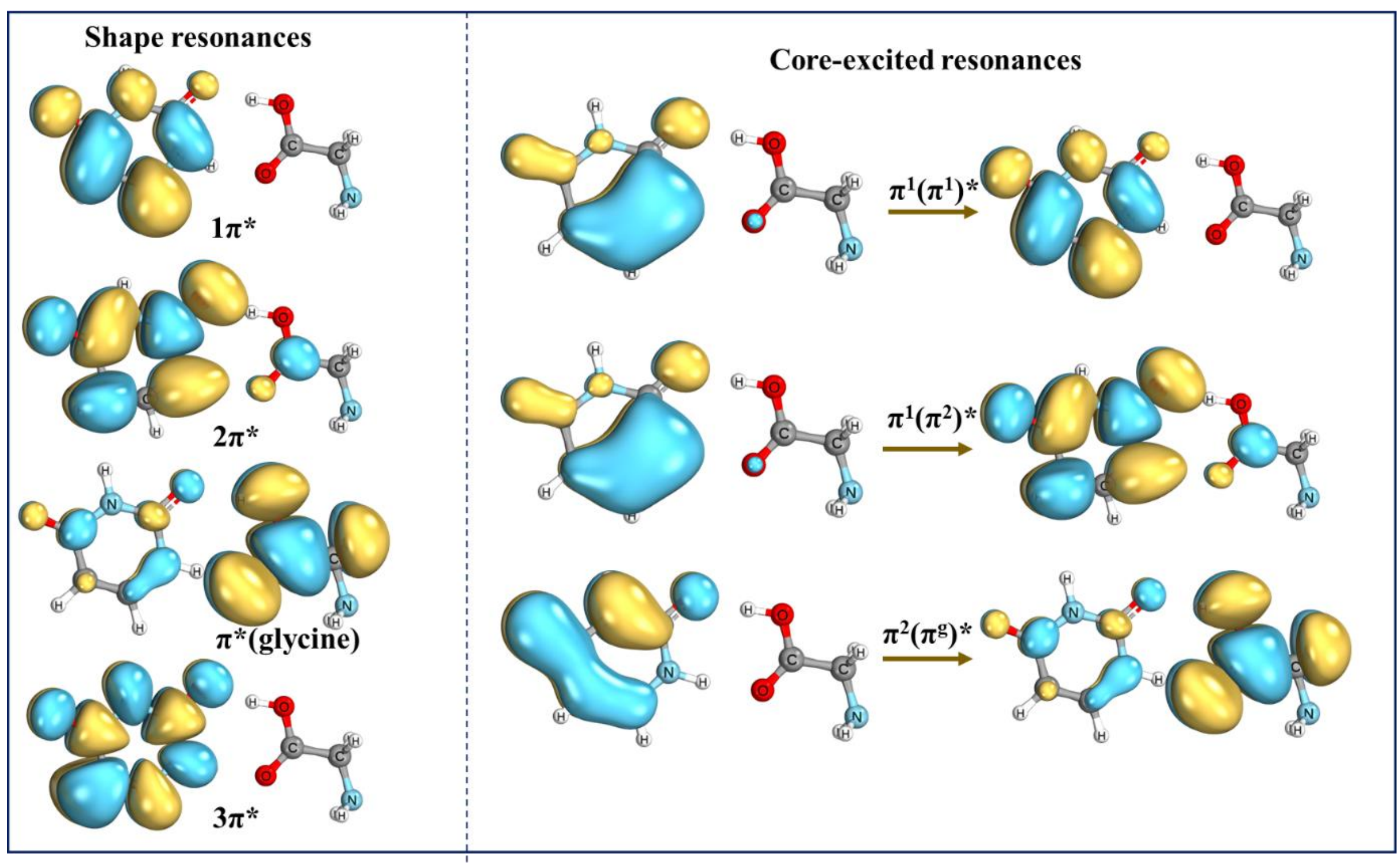

*Figure 7. Shape and core-excited resonances in the uracil(Gly) system in cc-pVDZ+1s1p1d basis with 1p scaling at the CASSCF/RVP level.*

*Table 1. The resonances energies Er and widths Γ (in parentheses) for isolated uracil anion obtained in this work at CAS(9,8)/cc-pVDZ+1s1p1d level, and comparisons with EA-EOM-DLPNO-CCSD/cc-pVDZ+2s2p2d and previously reported theoretical and experimental result.*

| | **1π*** | **2π*** | **3π*** | **$\pi^1(\pi^1)^*$** | **$\pi^1(\pi^2)^*$** |
|---|---|---|---|---|---|
| **CAS(9,8)** | 0.22(0.054) | 1.82(0.060) | 5.06(0.076) | 5.60(0.079) | 10.42(0.090) |
| **CAS(11,9)**[9] | 0.22(0.12) | 1.9(0.05) | 4.87(0.09) | 5.62(0.29) | - |
| **EA-EOM-DLPNO-CCSD**[35] | 0.22(0.014) | 1.54(0.050) | 4.42(0.121) | - | - |
| **Exp.**[36] | 0.22 | 1.58 | 3.83 | - | - |

*Table 2. The resonance positions and widths of shape and core-excited resonance states of uracil-glycine and uracil-ghost glycine calculated at the CAS/RVP level. (Values are in eV)*

| | **Energy** | **Width** | **Energy** | **Width** |
|---|---|---|---|---|
| **Shape** | **Uracil(Gly)** | | **Uracil(ghostGly)** | |
| 1π* | 0.22 | 0.044 | 0.22 | 0.070 |
| 2π* | 1.44 | 0.052 | 2.10 | 0.082 |
| π*(glycine) | 2.91 | 0.065 | | |
| 3π* | 5.04 | 0.070 | 5.23 | 0.101 |
| **Core-excited** | | | | |
| $\pi^{1}(\pi^{1})^{*}$ | 5.55 | 0.092 | 5.88 | 0.125 |
| $\pi^{1}(\pi^{2})^{*}$ | 8.60 | 0.117 | 10.85 | 0.152 |
| $\pi^{2}(\pi^{g})^{*}$ | 8.27 | 0.188 | | |